\documentclass[12pt]{article}

\usepackage[dvips]{graphics}

\usepackage[]{amsfonts}

\title{Superinflation and quintessence in the presence of an extra field}

\author{A.\ Figueiredo
\thanks{e-mail: annibal@helium.fis.unb.br}, 
T.\ M.\ Rocha Filho
\thanks{e-mail: marciano@helium.fis.unb.br}\\
{\em Departamento de F\'\i sica. Universidade de Bras\'\i lia. }\\
{\em 70.910-900 Bras\'\i lia-DF, Brazil }\\
E.\ Gunzig\\
{\em Instituts Internationaux de Chimie et de Physique Solvay.}\\
{\em RggR, Universit\'e Libre de Bruxelles.}\\
{\em CP 231, 1050 Bruxelles, Belgium}\\ 
L.\ Brenig\\
{\em Service de Physique Statistique, Universit\'e Libre de
Bruxelles.}\\
{\em CP231, 1050 Bruxelles, Belgium}\\
A.\ Saa\\
{\em Departamento de Matematica Aplicada, IMECC--UNICAMP.}\\ 
{\em CP6065, 13081-970 Campinas, SP, Brazil}}

\date{}

\begin{document}

\maketitle

\newpage
 
\abstract
We discuss the implication of the introduction of an extra field to the dynamics of a
scalar field conformally coupled to gravitation in a homogeneous isotropic spatially flat universe.
We show that for some reasonable parameter values the dynamical effects are similar to
those of our previous model with a single scalar field. Nevertheless for other parameter
values new dynamical effects are obtained.


\newpage

\section{Introduction}

Recently, the authors investigated~\cite{1Field} 
the dynamics of a spatially flat universe dominated by
a self-interacting conformally coupled scalar field according
to the action
\begin{equation}   \label{action1}
S=\frac{1}{2} \int d^4x\sqrt{-g}\left(-\, \frac{R}{\kappa } 
+g^{\mu \nu}\partial_\mu \psi \partial_\nu \psi 
-2V(\psi) +\xi R\psi^2  \right) \; ,
\end{equation}
where $R$ denotes the scalar  curvature, 
$\psi$ is the scalar fields,
$\kappa \equiv 8\pi G$ ($G$ being Newton's constant).
A cosmological constant, if
present, is incorporated in the scalar field potential 
\begin{equation}
V( \psi ) = \frac{3\alpha}{\kappa} \psi^2 -\frac{\Omega}{4}\psi^4 - 
\frac{9\omega}{\kappa^2} \;.
\end{equation}
Many new and interesting dynamical phenomena were discovered and its
cosmological interpretation were discussed. Special attention was given to the conformal
coupling case with $\xi=1/6$. This choice for the coupling constant is motivated by physical
arguments from particle theories~\cite{SonegoFaraoni} and from scale invariance at the classical
level~\cite{CCJ}. The non-minimal coupling is also required by first loop corrections~\cite{loop}.
One interesting effect discussed in~\cite{1Field} is superinflation characterized by $\dot H>0$, and that
can only be achieved by a non-minimal coupling.

Here we consider the
robustness of these dynamical phenomena with respect to the introduction into
the model of a second massless scalar field.
We also discuss some new type of asymptotic solutions
arising from the introduction of the extra field.
The structure of the paper is the following: in section 2 we present the model and some definitions. The fixed points
and their stability are given in section 3. Additional asymptotic solutions are discussed in section 4
and the paper is closed with some concluding remarks in section 5.

\section{The model}

We consider the conformally coupled theory described by the action:
\begin{eqnarray}   \label{action}
S & = & \frac{1}{2} \int d^4x\sqrt{-g}\left(-\, \frac{R}{\kappa } 
+g^{\mu \nu}\partial_\mu \psi_1 \partial_\nu \psi_1 
-2V_1(\psi_1)\right.\nonumber\\  & & \left.+\frac{1}{6} R\psi_1^2  
+g^{\mu \nu}\partial_\mu \psi_2 \partial_\nu \psi_2
-2V_2(\psi_2)
+ \frac{1}{6} R\psi_2^2 + \beta \psi_1^2\psi_2^2  \right).
\end{eqnarray}
We use the full conserved scalar field stress-energy tensor 
\begin{eqnarray}   \label{stressenergy}
T_{\mu\nu}  & = & \partial_{\mu}\psi_1 \partial_{\nu} \psi_1-
\frac{1}{6} \left(
\nabla_{\mu}\nabla_{\nu}
-g_{\mu\nu} \Box \right)( \psi_1^2 )+\frac{1}{6} G_{\mu\nu} \psi_1^2 
\nonumber \\
&-& \frac{1}{2} g_{\mu\nu} \left( \partial_{\alpha} \psi_1 
\partial^{\alpha} \psi_1 -2V_1( \psi_1) \right) \nonumber \\
&+& \partial_{\mu}\psi_2 \partial_{\nu} \psi_2-\frac{1}{6} \left(
\nabla_{\mu}\nabla_{\nu}
-g_{\mu\nu} \Box \right)( \psi_2^2 )+\frac{1}{6} G_{\mu\nu} \psi_2^2 
\nonumber \\
 &-& \frac{1}{2} g_{\mu\nu} \left( \partial_{\alpha} \psi_2 
\partial^{\alpha} \psi_2 -2V_2(\psi_2)\right) 
- \frac{\beta}{2}\psi_1^1\psi_2^2
\end{eqnarray}
(where $G_{\mu\nu}$ is the Einstein tensor), thereby 
avoiding the use of any effective coupling constant
in the Einstein equations.
Moreover, this consideration of the energy-momentum tensor $T_{\mu\nu}$ together with an
adequate self-consistent treatment of Einstein's and Klein-Gordon equations eliminates
the artificial pathologies associated, in Einstein's equations, with the ``critical factor''
$1-(\psi_1^2+\psi_2^2)\kappa/6$. The consensual attitude encountered in the literatures is
indeed that the dynamics is ill-defined when this factor is negative or vanishes. As we will see, that this is not
the case.

Let us consider the following form for the potential $V_1$ and $V_2$:
\begin{equation}  \label{potential}
V_i( \psi_i ) = \frac{3\alpha_i}{\kappa} \psi^2_i -\frac{\Omega_i}{4} \psi^4_i-\frac{9\omega_i}{\kappa^2}
 \;.
\end{equation}
The parameter $\alpha_i$ is related to the mass of the particle by $m_i=\sqrt{6\alpha_i/\kappa}$.
As in the first papers, we are interested in 
the dynamics of a spatially flat Friedmann-Robertson-Walker universe
with line element $ds^2=d\tau^2-a^2( \tau) \left(  dx^2+dy^2+dz^2
\right)$. Also we will restrict the potential in (\ref{potential}) to the case
$\alpha_1\equiv\alpha$, $\alpha_2=0$,
$\Omega_1\equiv\Omega$, $\Omega_2=0$, $\omega_1\equiv\omega$ and $\omega_2=0$.
This corresponds to a massive field coupled to a massless field.
The energy density and pressure associated to the scalar fields are:
\begin{equation}
\sigma=\sigma_1+\sigma_2-\frac{\beta}{2}\psi_1^2\psi_2^2,
\label{defsig}
\end{equation}
\begin{equation}
p=p_1+p_2+\frac{\beta}{2}\psi_1^2\psi_2^2,
\label{defp}
\end{equation}
where
\begin{equation}
\sigma_i=\frac{\dot\psi_i^2}{2}+\frac{1}{2}H^2\psi_i^2+\frac{1}{2}H\partial_\tau(\psi_i^2)+V_i(\psi_i),
\label{defsigi}
\end{equation}
\begin{equation}
p_i=\frac{\dot\psi_i^2}{2}-\frac{1}{6}\left[2H\partial_\tau(\psi_i^2)+\partial_{\tau\tau}^2(\psi_i^2)\right]
-\frac{1}{6}\left(2\dot H+2H^2\right)\psi_i^2-V_i(\psi_i).
\label{defpi}
\end{equation}

The action (\ref{action}) then implies the trace equation:
\begin{equation}
R=-6\left(\dot H+2H^2\right)=-\kappa \left(
\sigma-3p\right),
\label{treq}
\end{equation}
the energy constraint:
\begin{equation}
3H^2-\kappa\sigma=0,
\label{enconteq}
\end{equation}
and the Klein-Gordon equations for the two scalar fields:
\begin{equation}
\ddot\psi_1+3H\dot\psi_1-\frac{1}{6}R\psi_1+\frac{dV_1}{d\psi_1}+\beta\psi_1\psi_2^2=0,
\label{klein1}
\end{equation}
\begin{equation}
\ddot\psi_2+3H\dot\psi_2-\frac{1}{6}R\psi_2+\frac{dV_2}{d\psi_2}+\beta\psi_1^2\psi_2=0.
\label{klein2}
\end{equation}

Using the Klein-Gordon equations (\ref{klein1}) and (\ref{klein2}) and the expressions for $\sigma$ and $p$
we obtain
\begin{equation}
\sigma-3p=4\left(V_1+V_2\right)-\psi_1\frac{dV_1}{d\psi_1}-\psi_2\frac{dV_2}{d\psi_2}.
\label{sigm3p}
\end{equation}
This simple form holds for any interaction of the form $\beta\psi_1^\gamma\psi_2^\delta$ provided
$\gamma+\delta=4$.
Finally, the energy constraint is:
\begin{eqnarray}
\lefteqn{3H^2-\kappa\sigma=3H^2-\frac{1}{2}\kappa\dot\psi_1^2-\frac{1}{2}\kappa H^2\psi_1^2-\kappa H\psi_1\dot\psi_1
-\kappa V_1\nonumber}\\
 & & -\frac{1}{2}\kappa\dot\psi_2^2-\frac{1}{2}\kappa H^2\psi_2^2-\kappa H\psi_2\dot\psi_2-\kappa V_2
+\frac{1}{2}\kappa\beta\psi_1^2\psi_2^2=0.
\label{eq12}
\end{eqnarray}
Now solving eqs.\ (\ref{sigm3p}) and (\ref{eq12}) for the derivatives $\dot H$ and $\dot\psi_1$ we obtain:
\begin{equation}
\dot H=-2H^2+\frac{2}{3}\kappa V_1+\frac{2}{3}\kappa V_2-\frac{1}{6}\kappa\psi_1\frac{dV_1}{d\psi_1}
-\frac{1}{6}\kappa\psi_2\frac{dV_2}{d\psi_2},
\label{eq15}
\end{equation}
\begin{equation}
\dot\psi_1=\frac{1}{\kappa}\left(-\kappa H\psi_1
\pm\sqrt{G}\right),
\label{eq16}
\end{equation}
where
\begin{equation}
G\equiv 6\kappa H^2-\kappa^2\dot\psi_2^2-2\kappa^2 H\psi_2\dot\psi_2
-2\kappa^2(V_1+V_2)-\kappa^2\beta\psi_1^2\psi_2^2-\kappa^2 H^2\psi_2^2.
\label{defg}
\end{equation}
The region $G<0$ in phase space is physically forbidden.

Using eq.\ (\ref{eq16}) allows to rewrite system (\ref{defsig}--\ref{treq}) in the following equivalent form:
\begin{equation}
\dot H=-2H^2+\frac{\kappa(\sigma-3p)}{6},
\label{sys4_1}
\end{equation}
\begin{equation}
\dot\psi_1=-H\psi_1\pm\frac{\sqrt{G}}{\kappa},
\label{sys4_2}
\end{equation}
\begin{equation}
\ddot \psi_2=-3H\dot\psi_2+\frac{1}{6}R\psi_2-\frac{dV_2}{d\psi_2}-\beta\psi_1^2\psi_2.
\label{sys4_3}
\end{equation}
The system of ODE (\ref{treq},\ref{klein1},\ref{klein2})
is a five-dimensional system, but due to energy constraint
(\ref{enconteq}) the real phase space is constrained
into a four-dimensional manifold. The situation is similar to the
one field case, where the original system is three dimensional, but
the energy constraint enforces the motion to take place on a
two dimensional manifold. In the one field case, the plane
$(H,\psi)$ is divided in sectors by some straight lines, corresponding
to regions of distinct state equations for the fluid $\psi$.
In the present case, the projection of the real phase-space onto
the $(H,\psi_1)$ plane conserves some of that sectors. For example,
{}from
\begin{equation}
\sigma-3p = 6\alpha\psi_1^2,
\end{equation}
valid for any quartic interaction potential between $\psi_1$ and
$\psi_2$, one can see that the region $\psi_1=0$ still corresponds
to a state equation of pure radiation.

\section{Fixed points}

In our case De Sitter solutions correspond to fixed points except for the fixed point at the origin.
The fixed points are obtained from the conditions $\dot\psi_1=\dot\psi_2=0$. The stability of the fixed
points is determined from the linearized equations in a neighborhood of each point, except for special cases discussed below.
For all the fixed points we have, in the formula below,
 $s_1,s_2,s_3=\pm 1$.
The existence conditions for the fixed points is such
that the arguments of the square roots are positive:

\begin{itemize}

\item{Type A}

\begin{equation}
H=s_1\sqrt{\frac{3}{\kappa}}\sqrt{\frac{-\beta}{\alpha+\beta}},
\label{pf2a_1}
\end{equation}
\begin{equation}
\psi_1=s_2\sqrt{\frac{6}{\kappa}}\sqrt{\frac{\omega}{\alpha+\beta}},
\label{pf2a_2}
\end{equation}
\begin{equation}
\psi_2=s_3\sqrt{\frac{6}{\kappa}}\sqrt{\frac{\omega(\Omega+\beta)-\alpha(\alpha+\beta)}
{\beta(\alpha+\beta)}}.
\end{equation}

\item{Type B}
\begin{equation}
H=s_1\sqrt{\frac{-3\omega}{\kappa}},
\label{pf2b_1}
\end{equation}
\begin{equation}
\psi_1=\psi_2=0.
\label{pf2b_2}
\end{equation}

\item{Type C}
\begin{equation}
H=s_1\sqrt{\frac{3}{\kappa}}\sqrt{\frac{\alpha^2-\omega\Omega}{\Omega-\alpha}},
\label{pf2c_1}
\end{equation}
\begin{equation}
\psi_1=s_2\sqrt{\frac{6}{\kappa}}\sqrt{\frac{\alpha-\omega}{\Omega-\alpha}},
\label{pf2c_2}
\end{equation}
\begin{equation}
\psi_2=0.
\label{pf2c_3}
\end{equation}
Fixed points of type B are on the $G=0$ surface.

\end{itemize}

In order to discuss the stability of these fixed points
let us consider the generic case given by a system of the form
\begin{equation}
\dot x_i=F_i(x_1,\ldots,x_n)\equiv F(x),\hspace{10mm}i=1,\ldots,n.
\label{gen1}
\end{equation}
A fixed point $\overline{x}$ satisfies $F_i(\overline{x})=0$. It stability is usually established by
linearizing (\ref{gen1}) in a neighborhood of $\overline{x}$. In this way we write $x=\overline{x}+\delta x$
and obtain (\ref{gen1}):
\begin{equation}
\frac{d\:\delta x}{dt}={\bf J}\delta x+{\cal O}(\delta x^2),
\label{gen2}
\end{equation}
where ${\bf J}$ is defined by
\begin{equation}
{\bf J}_{ij}\equiv\left.\frac{\partial F_i}{\partial x_j}\right|_{x=\overline{x}},
\label{gen3}
\end{equation}
and $\delta x\equiv(\delta x_1,\ldots,\delta x_n)$. If $Det({\bf J})\neq0$ and if the eigenvalues
$\lambda_i$ of ${\bf J}$ are such that ${\rm Re}(\lambda_i)\neq0$ then the Hartman-Grobmann theorem ensures that
the linearized system obtained by retaining only linear terms in (\ref{gen2}) is topologically equivalent
to the original system (\ref{gen1}) in a neighborhood of the fixed point.
The real parts of the eigenvalues then determine the local stability
of the fixed points. The stability of the fixed points on the $G=0$ surface cannot be studied using this method
as $Det({\bf J})=0$. To overcome this problem we will consider the system formed by eqs.\ (\ref{klein1}), (\ref{klein2})
and (\ref{eq15}). The only points to be considered separately are of type B and their
stability is studied by other methods.

Equations (\ref{klein1}) and (\ref{klein2}) can be rewritten as a first order equations by introducing
the new variables $\phi_1\equiv\dot\psi_1$ and $\phi_2\equiv\dot\psi_2$. In this way we identify the
variables $x_i$ in (\ref{gen1}) with $(H,\psi_1,\phi_1,\psi_2,\phi_2)$ and obtain from (\ref{gen3}) the
following form for the ${\bf J}$ matrix computed at the fixed point of coordinates
$(\overline{H},\overline{\psi}_1,\overline{\phi}_1,\overline{\psi}_2,\overline{\phi}_2)$:
\begin{equation}
{\bf J}=\left(
\begin{array}{ccccc}
-4\overline{H} & 2\alpha\overline{\psi_1} & 0 & 0 & 0\\
0 & 0 & 1 & 0 & 0\\
0 & C_1 & -3\overline{H} & -2\beta\overline{\psi}_1\overline{\psi}_2 & 0\\
0 & 0 & 0 & 0 & 1\\
0 & -2(\alpha+\beta)\overline{\psi}_1\overline{\psi}_2 & 0 & C_2 & -3\overline{H}
\end{array}
\right),
\label{gen4}
\end{equation}
with
\begin{equation}
C_1=-\beta\overline{\psi}_2^2+3(\Omega-\alpha)\overline{\psi}_1^2+\frac{6}{\kappa}(\omega-\alpha),
\label{gen5}
\end{equation}
and
\begin{equation}
C_2=-(\alpha+\beta)\overline{\psi}_1^2+6\frac{\omega}{\kappa}.
\label{gen6}
\end{equation}
The eigenvalues of ${\bf J}$ are:
\begin{equation}
\lambda=-\frac{3}{2}\overline{H}\pm\frac{1}{2}\sqrt{9\overline{H}+9A\pm\frac{2}{\kappa}\sqrt{B}},
\label{gen7}
\end{equation}
\begin{equation}
\lambda=-4\overline{H},
\label{geb7}
\end{equation}
where the $\pm$ signs are to be considered independently of each other, 
\begin{equation}
A=\frac{2}{9}(3\Omega-\beta-4\alpha)\overline{\psi}_1^2-\frac{2}{9}(\beta+4\alpha)\overline{\psi}_2^2
+\frac{4}{3\kappa}(2\omega-\alpha),
\label{gen11}
\end{equation}
and
\begin{eqnarray}
\lefteqn{B=\kappa^2(\beta^2\overline{\psi}_2^2
+(20\alpha\beta+14\beta^2-6\beta\Omega)
\overline{\psi}_1^2\overline{\psi}_2^2}\nonumber\\
 & & +(4\alpha^2+\beta^2-4\alpha\beta-12\Omega\alpha+6\beta\Omega+9\Omega^2)\overline{\psi}_1^4)\;.
\label{gen12}
\end{eqnarray}
A fixed point is stable, unstable or a saddle point if the real parts of the five eigenvalues are all negative,
all positive or have different signs, respectively. Points of Type A and C are saddle points.
The stability of fixed points of type B is discussed in the next section

\section{Asymptotic dynamics}

\begin{figure}
%\begin{center}
\scalebox{0.25}{{\includegraphics{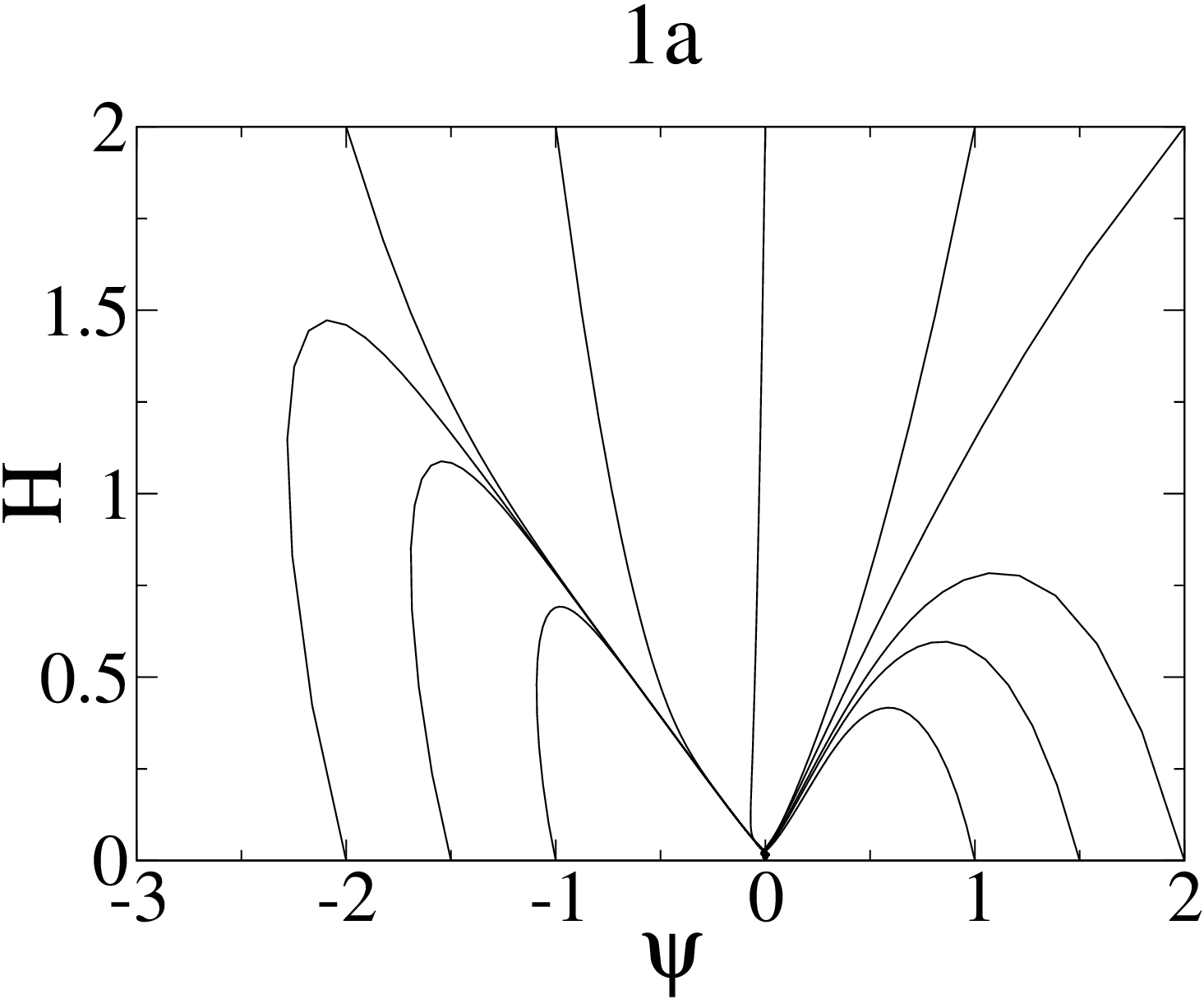}}}\hspace{10mm}
\scalebox{0.25}{{\includegraphics{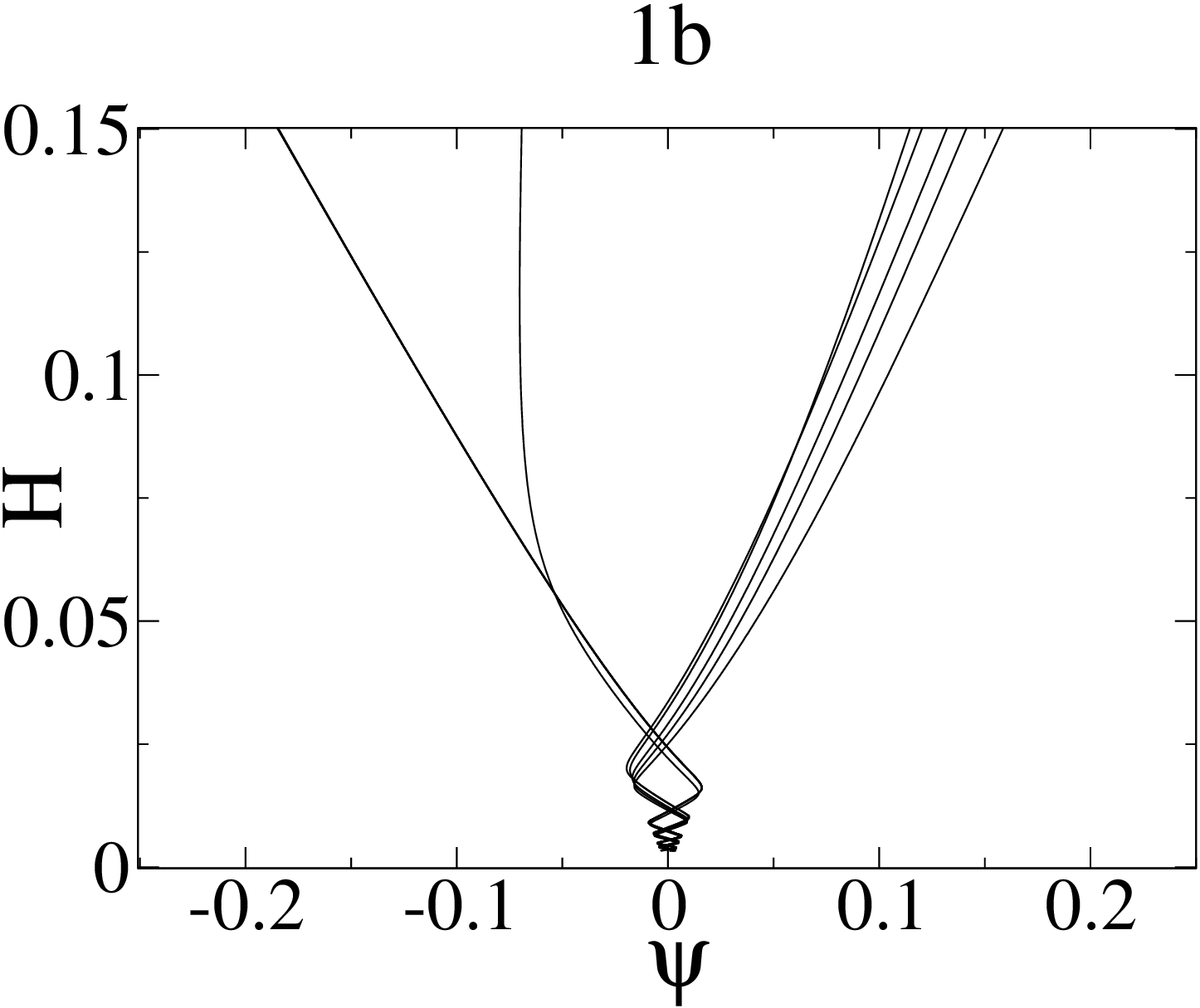}}}
%\end{center}
\caption{Example of solutions projected in the $(H,\psi_1)$ plane for $\omega=0$. The parameter values used are
$\alpha=1$, $\Omega=1/2$, $m=1$, $\beta=-1.2345$. The solutions are attracted to the fixed point at the origin.}
\label{fig1}
\end{figure}

\begin{figure}
\scalebox{0.25}{{\includegraphics{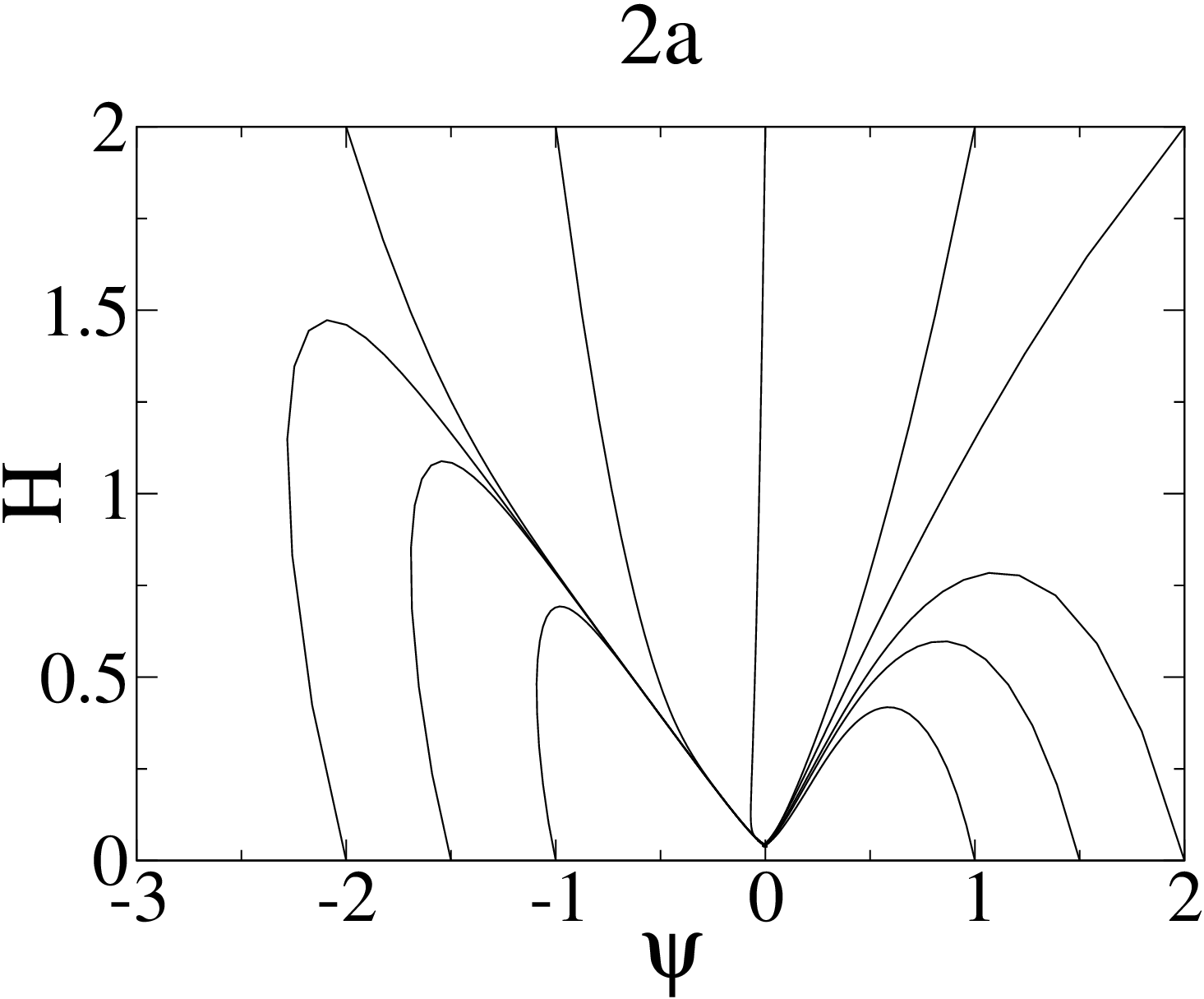}}}\hspace{10mm}
\scalebox{0.25}{{\includegraphics{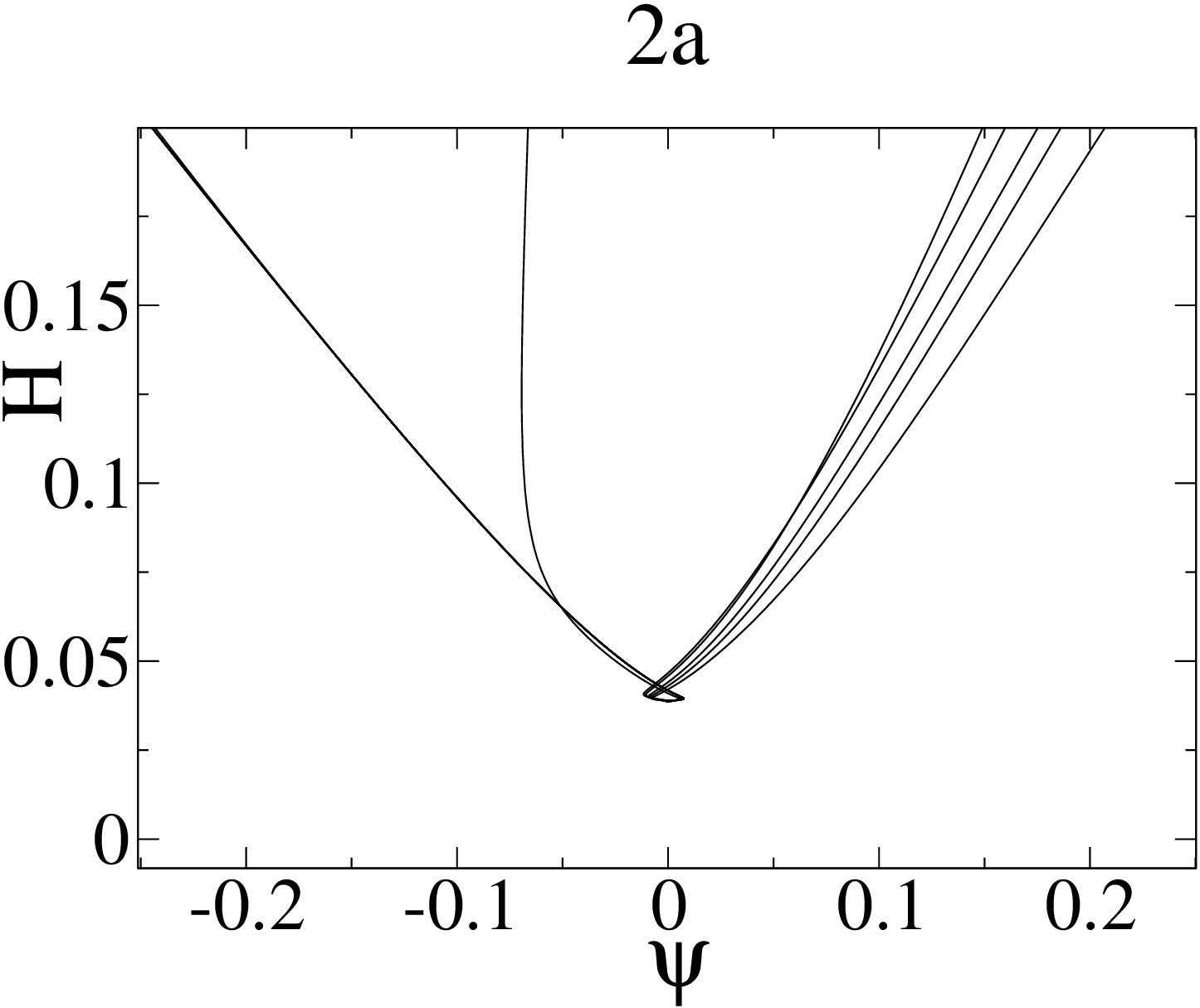}}}
\caption{Example of solutions projected in the $(H,\psi_1)$ plane for $\omega=-0.5$. The other parameter values are
the same as in fig.~\ref{fig1}. The solutions are attracted to the fixed point at $\psi_1=0$ and $H=\sqrt{-3\omega/\kappa}$.}
\label{fig2}
\end{figure}

Two kinds of asymptotic dynamics are more relevant for our model: attractive fixed points and diverging solutions.
For the one field model the fixed point at $H=\psi=0$ acts as a an attractor and plays an important role in our approach~\cite{1Field}.
The fixed point at
\begin{equation}
\psi_1=\dot\psi_1=\psi_2=\dot\psi_2=0 \hspace{2mm}{\rm and}\hspace{2mm} H=\sqrt{-3\omega/\kappa},
\label{fp}
\end{equation}
has a similar role.
For this purpose we will consider here the special case of a non-negative cosmological constant $\omega\leq0$.
This implies that $\dot H\geq0$ for $H=0$ as $\sigma-3p$ is positive for $\omega$ negative and therefore the region
$H>0$ is invariant under the dynamics. Let us now consider the function
\begin{equation}
L=\frac{1}{2}\dot\psi_1^2+\frac{1}{2}\frac{\beta}{\beta-\alpha}\dot\psi_2^2+\frac{3\alpha}{\kappa}\psi_1^2
+\frac{1}{4}(\alpha-\Omega)\psi_1^4-\frac{1}{2}\beta\psi_1^2\psi_2^2.
\label{lyap1}
\end{equation}
Its total time derivative modulo eqs.\ (\ref{klein1}), (\ref{klein2}) and (\ref{eq15}) is given by
\begin{equation}
\dot L=-3H\left(\dot\psi_1^2+\frac{\beta}{\beta-\alpha}\dot\psi_2^2\right)+\frac{6\omega}{\kappa}\psi_1\dot\psi_1
+\frac{6\beta\omega}{(\beta-\alpha)\kappa}\psi_2\dot\psi_2.
\label{lyap2}
\end{equation}
For parameter values such that $\omega=0$, $\beta<0$ and $\Omega\leq\alpha$ the function $L$ is a Lyapunov function
for the fixed point (\ref{fp}) and satisfy the properties
\begin{equation}
L\geq0,
\label{lyap3}
\end{equation}
\begin{equation}
\dot L\leq0.
\label{lyap4}
\end{equation}
The equality in (\ref{lyap3}) and (\ref{lyap4}) is valid only on the fixed point. Hence the fixed point is a global attractor
for the whole region $H\geq0$ (see fig.~\ref{fig1}).
Numerical simulations indicate that the same is also true for $\omega<0$ as shown in fig.~\ref{fig2}.

For other parameter values numerical solutions point for an asymptotic diverging solution
(see fig.\ \ref{fig3}). For simplicity, let us consider the case $\omega=0$ and the
ansatz:
\begin{figure}
\scalebox{0.25}{{\includegraphics{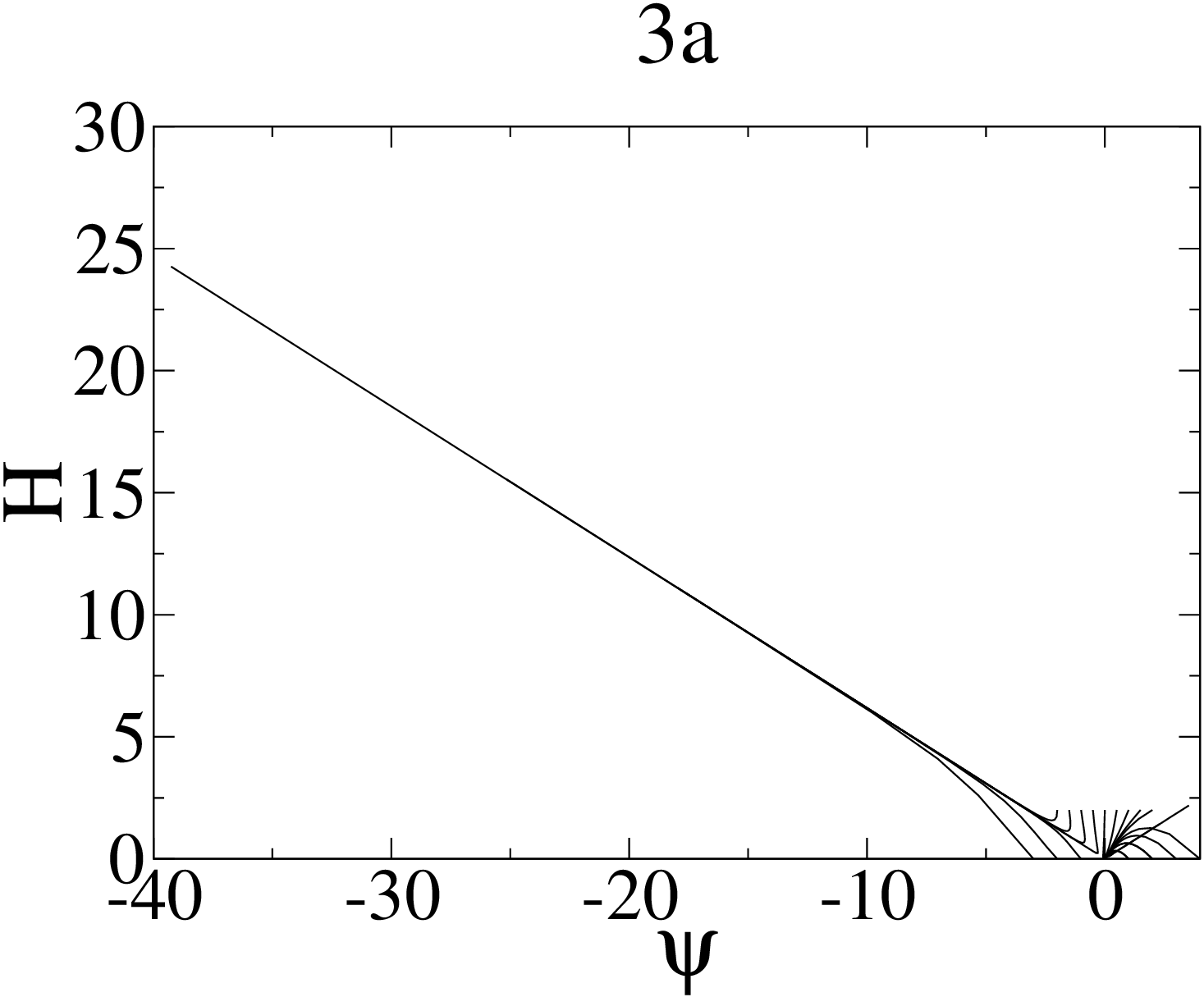}}}\hspace{10mm}
\scalebox{0.25}{{\includegraphics{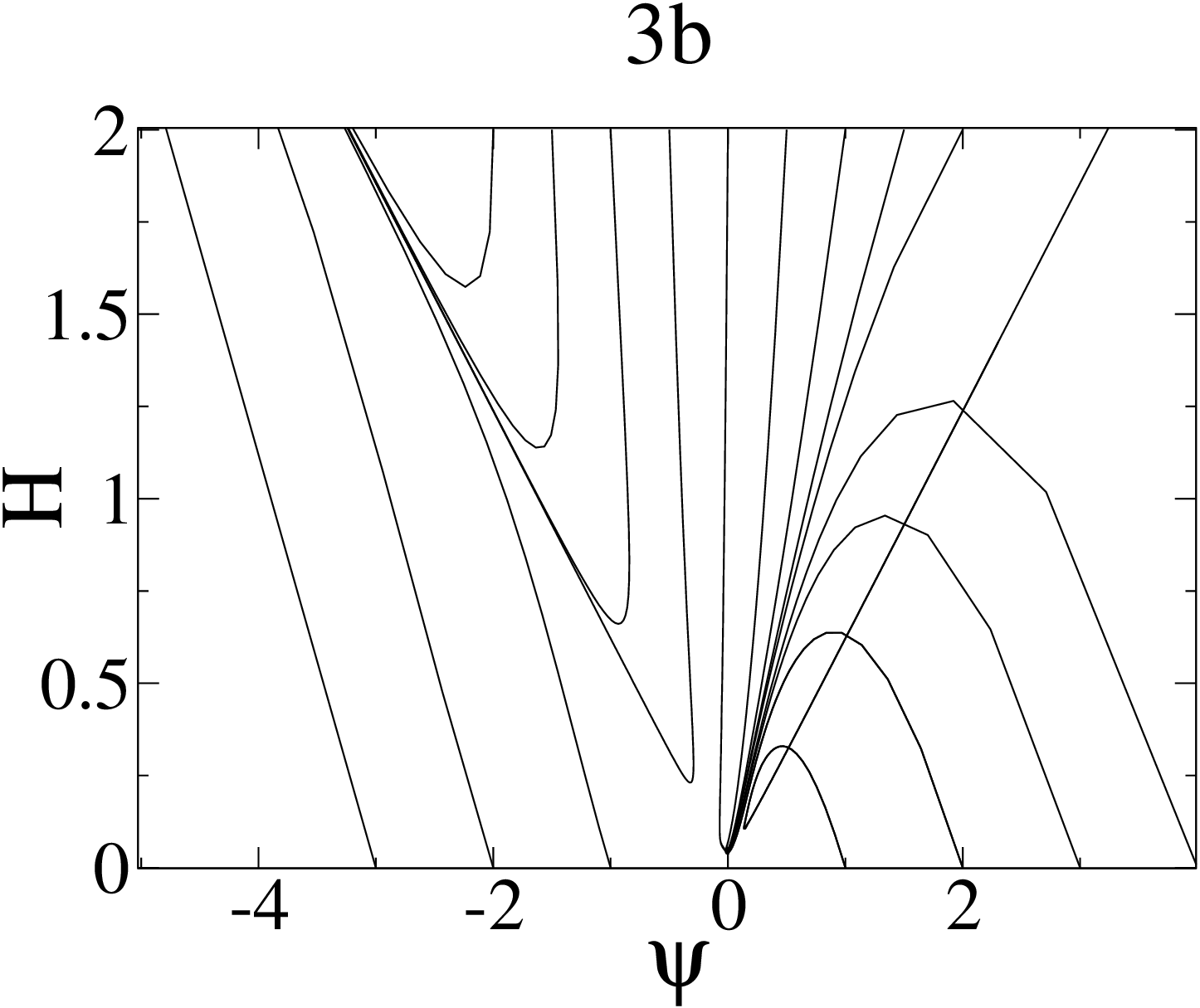}}}
\label{fig3}
\caption{Solution with $\omega=-0.3$. The remaining parameters are as in fig.~\ref{fig1}. Some solutions are diverging
while ore are still attracted to the fixed point at the origin. The quantities $H$, $\psi_1$ and $\psi_2$ diverge in a finite time.}
\end{figure}
\begin{equation}
H=\delta\psi_1,
\label{eq17}
\end{equation}
with $\delta$ a constant. Hence
\begin{equation}
\dot H=\delta\dot\psi_1.
\label{eq18}
\end{equation}
From eqs.\ (\ref{eq15}), (\ref{eq16}) and (\ref{eq18}) we obtain
\begin{equation}
\delta\dot\psi_1=\frac{1}{6}\psi_1^2(-12\delta^2+m_1^2\kappa),
\label{eq19}
\end{equation}
which is a closed equation for $\psi_1$, with solution
\begin{equation}
\psi_1(t)=\frac{6\delta\psi_1(0)}{6\delta+(12\delta^2-\kappa m_1^2)\psi_1(0)t},
\label{eq20}
\end{equation}
where $\psi_1(0)$ is the initial condition for $\psi_1$ at $t=0$. Using eq.\ (\ref{klein1}) and solution
(\ref{eq20}) yields easily $\psi_2(t)$:
\begin{equation}
\psi_2(t)=\pm\sqrt{-\frac{2}{\beta}}\frac{\sqrt{36\delta^4-12\delta^2m_1^2\kappa-18\Omega\delta^2+m_1^4\kappa^2}\:\psi_1(0)}
{(12\delta^2-\kappa m_1^2)\psi_1(0)t+6\delta}.
\label{eq21}
\end{equation}
 
With the expressions for $\psi_1(t)$ and $\psi_2(t)$ it is  straightforward (but a little bit cumbersome) to
obtain the proportionality constant $\delta$ from the remaining equation (\ref{eq16}):
\begin{equation}
\delta=\pm\frac{1}{6}\sqrt{6\kappa m_1^2-9\beta+3\sqrt{-12\beta\kappa m_1^2+9\beta^2}}.
\end{equation}
This value for $\delta$ agrees with the results of numerical integrations. It is important to note at this point
that not necessarily every solution has the asymptotic behavior of eq.\ (\ref{eq17}). From numerical investigations
it seems to exist a threshold in the initial values of $\psi_2$ and $\dot\psi_2$ such that above it the solutions converge
asymptotically to the behavior obtained above. Below this threshold solutions near the origin in the $(H,\psi_1)$ plane
are attracted towards it, exhibiting a similar behavior to the case of a single massive scalar field.

\section{Concluding remarks}

The introduction of an extra massless field conserve some essential features of the dynamics of the
one-field model of ref.~\cite{1Field}, for some parameter values, particularly the spiraling
solution near the fixed point at the origin. Nevertheless some new type of solutions
exist that have no analog in the one-field case. For two self-interacting coupled fields
the behavior is even more richer and a more exhaustive study is under preparation.
Nevertheless, for some parameter ranges, new dynamical effects are obtained which modify the model
in a substantial way. A discussion of which parameter values are of physical relevance is important
for the future development of the present approach.

\section{Acknowledgments}

We acknowledge the Centre de Calcul Symbolique sur Ordinateur (Brussels) for
the use of computer facilities, and financial support from the EEC
(grant HPHA-CT-2000-00015), from OLAM, Fondation pour la Recherche
Fondamentale (Brussels), from CNPq and FAPESP (Brazil), and from the
ESPRIT Working Group CATHODE.

\end{document}